\documentclass[12pt]{JHEP3}

\usepackage{graphicx}
\usepackage{psfrag}
\usepackage{axodraw}
\usepackage{amsmath}

\def\beq{\begin{equation}}
\def\eeq{\end{equation}}
\def\bea{\begin{eqnarray}}
\def\eea{\end{eqnarray}}
\def\nn{\nonumber}

\def\bra#1{\langle #1|}
\def\ket#1{| #1\rangle}
\def\roughly#1{\mathrel{\raise.3ex\hbox
{$#1$\kern-.75em\lower1ex\hbox{$\sim$}}}}
\def\lsim{\roughly<}

\def\sss{\scriptscriptstyle}

\def\bd{B_d^0}

\def\bs{B_s^0}
\def\bsbar{{\bar B}_s^0}

\def\kbar{{\bar K}^0}

\def\btod{{\bar b} \to {\bar d}}
\def\btos{{\bar b} \to {\bar s}}

\def\ANPu{{\cal A}^u}
\def\ANPd{{\cal A}^d}

\def\mw{M_{\sss W}}

\def\bskk{{\bs}\to K^+ K^-}
\def\bskkneut{{\bs}\to K^0 \kbar}
\def\bdpipi{{\bd}\to \pi^+ \pi^-}
\def\bdKK{{\bd}\to K^0 \kbar}

\def\ANPu{{\cal A}^u}
\def\ANPd{{\cal A}^d}

\title{\boldmath $\bskk$ and $\bskkneut$ Decays within Supersymmetry}

\author{Seungwon Baek\\
Department of Physics, Yonsei University, Seoul 120-749, Korea \\
E-mail: \email{swbaek@kias.re.kr}}

\author{David London\\
Physique des Particules, Universit\'e de Montr\'eal, \\
C.P. 6128, succ.~centre-ville, Montr\'eal, QC, Canada H3C 3J7 \\
E-mail: \email{london@lps.umontreal.ca}}

\author{Joaquim Matias\\
IFAE, Universitat Aut\`onoma de Barcelona, \\
08193 Bellaterra, Barcelona, Spain \\
E-mail: \email{matias@ifae.es}}

\author{Javier Virto\\
IFAE, Universitat Aut\`onoma de Barcelona, \\
08193 Bellaterra, Barcelona, Spain \\
E-mail: \email{jvirto@ifae.es}}

\abstract{ We compute the supersymmetric (SUSY) contributions to the
observables in $\bskk$ and $\bskkneut$ decays. The hadronic parameters
in the standard-model (SM) amplitudes are obtained from the $\bdKK$
decay using a recent approach that combines flavor SU(3) symmetry and
a controlled input from QCD factorization. The latest experimental
data for $BR(\bskk)$ is in agreement with the SM prediction. We study
how the branching ratios and the direct and mixing-induced CP
asymmetries of both $\bs \to K{\bar K}$ decay modes are affected with
the inclusion of SUSY, after imposing constraints from $BR(B \to X_s
\gamma)$, $B \to \pi K$ and $\Delta M_s$ over the parameter space.
While the branching ratios remain unaffected by SUSY, we identify the
CP asymmetries of the $\bs \to K{\bar K}$ decays as the most promising
observables to look for large deviations from the SM. }


\preprint{UdeM-GPP-TH-06-150  \\ UAB-FT-613 }

\begin{document}

\section{Introduction}

Recently, the precision of $B$-physics measurements has increased
dramatically due to the experimental results of CDF, Babar and
Belle. This implies that new strategies are necessary for
controlling hadronic uncertainties. In addition, it is important to
identify those observables which are useful for signalling the
presence of physics beyond the standard model (SM).  Once these are
found, the next step is to explore the impact of well-motivated
models.

In a recent publication \cite{BLMV} we computed the supersymmetric
(SUSY) contributions to $\bskk$ decays. In the present paper, we
present an update of this analysis, with four important
improvements. First, we extend the calculation to include
$\bskkneut$ decays. Second, our discussion of the predictions of the
SM is based on a new method which uses the $\bdKK$ decay \cite{DMV}.
It combines QCD factorization with flavour symmetries, and
represents a substantial improvement in the control of hadronic
uncertainties. Third, the limits on the SUSY parameter space include
the latest constraints from $\bs$--$\bsbar$ mixing \cite{Bsmix}, as
well as data from $B\to X_s \gamma$ and $B\to\pi K$ decays. The
fourth point is related to squark mixing.  For each fermion, each of
the two components (left-handed, right-handed) has a scalar SUSY
partner, the squark. One can have mixing of the left-handed or
right-handed squarks of the three generations. In Ref.~\cite{BLMV},
it was assumed that one of the two mixings (LL or RR) was zero.  In
this paper, we allow simultaneous nonzero values of both LL and RR
mixing.

We pay particular attention here to the CP-violating asymmetries of
the decays $\bskk$ and $\bskkneut$. These are defined as:
\beq A_{dir} = { \left\vert {\cal A} \right\vert^2 - \left\vert
{\bar{\cal A}} \right\vert^2 \over \left\vert {\cal A} \right\vert^2
+ \left\vert {\bar{\cal A}} \right\vert^2} ~~,~~~~ A_{mix} = -2\ {
{\rm Im} \left( e^{-i\phi_s} {\cal A}^* {\bar{\cal A}} \right) \over
\left\vert {\cal A} \right\vert^2 + \left\vert {\bar{\cal A}}
\right\vert^2} ~, \eeq
where $\cal A$ is the amplitude of the decay in question, $\bar{\cal
A}$ is the amplitude of the CP-conjugate process, and $\phi_s$ is
the phase of $\bs$--$\bsbar$ mixing.

In the SM, $\bskkneut$ is dominated by a single penguin decay
amplitude. There are other penguin contributions, but they are
suppressed and enter only at the level of $\le 5\%$. (They are all
included in Sec.~2.) The direct CP-violating asymmetry, $A_{dir}$,
involves the interference of the dominant penguin amplitude and the
suppressed contributions, and is therefore very small in the SM. As
for the mixing-induced CP asymmetry, $A_{mix}$, since the
$\bskkneut$ decay is dominated by a single amplitude, $A_{mix}$
essentially measures $\phi_s$. This phase is very small in the SM
\cite{pdg}, so that this CP asymmetry is expected to be
correspondingly small. Since both CP asymmetries of $\bskkneut$ are
expected to be so small in the SM, this makes them interesting
observables for detecting the presence of new physics (NP).  The
situation is somewhat different for the decay $\bskk$ since it
receives a tree contribution which cannot be neglected with respect
to the dominant penguin contribution. In the SM, the interference of
the penguin and tree amplitudes in $\bskk$ gives rise to larger CP
asymmetries than in $\bskkneut$. Both decays involve a $\btos$
transition and therefore have branching ratios of $O(10^{-5})$.

All of these SM predictions can change in the presence of SUSY.
Naively, one would guess that all SUSY contributions to $\bskk$ and
$\bskkneut$ are suppressed by $\mw^2/M_{\sss SUSY}^2$, where
$M_{\sss SUSY} \sim 1$ TeV, and are therefore small. However, some
of the SUSY contributions involve squark-gluino loops, which are
proportional to the strong coupling constant $\alpha_s$.  Compared
to the SM, the relative size of these contributions is therefore
$(\alpha_s/\alpha)(\mw^2/M_{\sss SUSY}^2)$. Since this is $O(1)$,
such contributions can compete with those of the SM, leading to
significant modifications of the SM predictions for $\bskk$, and
especially for $\bskkneut$. In this paper, we consider only these
SUSY contributions, as they are the dominant effects.

Experimentally, the branching ratio of $\bskk$ has been measured at
CDF\cite{CDF}:
\beq BR(\bskk)_{\rm exp}=(24.4\pm 1.4 \pm 4.6) \times 10^{-6} ~.
\label{bskk_CDF} \eeq
The CP asymmetries for $\bskk$ are likely to be measured soon at CDF
\cite{Punzi}. However, the measurements for $\bskkneut$ will
probably take more time at CDF or may have to wait until LHCb.

In Sec.~2, we discuss the SM expectations for the various
observables in $\bskk$ and $\bskkneut$. In order to do this, we must
consider certain $\bd$ decays. Here we follow Ref.~\cite{DMV} and
use $\bdKK$. Sec.~3 contains the general analysis of $\bskk$ and
$\bskkneut$ decays with the addition of NP. We discuss the
amplitudes for these decays in the presence of NP, as well as
strategies for measuring the NP parameters. We turn specifically to
SUSY in Sec.~4 and calculate its effect on the amplitudes of $\bskk$
and $\bskkneut$. We note that SUSY can substantially modify the SM
predictions for the CP asymmetries while keeping the branching
ratios basically unaffected. We conclude in Sec.~5.

\section{\boldmath Standard-Model Analysis of $\bskk$ and $\bskkneut$}

Consider first $\bskk$, which at the quark level is ${\bar b} \to
{\bar s} u {\bar u}$. This decay receives a contribution from a
penguin diagram ${\scriptstyle PEN}'$ (the prime indicates a $\btos$
transition). The penguin diagram receives contributions from each of
the internal quarks $u$, $c$ and $t$. However, using the unitarity
of the Cabibbo-Kobayashi-Maskawa (CKM) matrix, one can write
\beq {\scriptstyle PEN}' = V_{ub}^* V_{us} (P'_u - P'_t) + V_{cb}^*
V_{cs} (P'_c - P'_t) ~. \eeq
The decay $\bskk$ also receives several other diagrammatic
contributions \cite{GHLR}. The most important of these is the tree
amplitude, which is proportional to $V_{ub}^* V_{us}$:
${\scriptstyle TREE}'=V_{ub}^* V_{us}\,T'$. The amplitude can
therefore be written\footnote{We use the following notation (see
Ref.~\cite{DMV}): quantities carrying the superscripts $d0$, $s0$
and $s\pm$ correspond to $\bdKK$, $\bskkneut$ and $\bskk$,
respectively.}
\bea {\cal A}(\bs \to K^+ K^-) & \simeq & V_{ub}^* V_{us}[T'+
  (P'_u - P'_t)] + V_{cb}^* V_{cs} (P'_c - P'_t) \nn\\
& \equiv & V_{ub}^* V_{us} T^{s\pm} + V_{cb}^* V_{cs} P^{s\pm} ~,
\eea
where $P^{s\pm} \equiv (P'_c - P'_t)$ and $T^{s\pm} = [T' + (P'_u -
P'_t)]$. Note that $|V_{ub}^* V_{us}| \simeq 5\% |V_{cb}^* V_{cs}|$,
and this CKM suppression compensates the relative size of the
amplitudes $|P^{s\pm}/T^{s\pm}|\sim 0.1$ (see Ref.~\cite{DMV}).
Thus, the first term is smaller than the second, but must be
included in the analysis.

The amplitude for $\bskkneut$ (quark level: ${\bar b} \to {\bar s} d
{\bar d}$) can be treated similarly. In this case, there is no tree
diagram, but we keep the notation $T^{s0}$ for the penguin
contribution proportional to $V_{ub}^* V_{us}$, leading to
\beq {\cal A}(\bskkneut) \simeq V_{ub}^* V_{us} T^{s0} + V_{cb}^*
V_{cs} P^{s0} ~. \eeq
Here, both $T^{s0}$ and $P^{s0}$ are of the same size, but since
nothing compensates for the strong CKM suppression, the first term
is strongly suppressed, leading to a very small direct CP asymmetry.

In the isospin limit, $P^{s\pm} = P^{s0}$. However, $T^{s\pm} \ne
T^{s0}$ due to the presence of the tree diagram in $\bskk$. Since
the terms proportional to $V_{ub}^* V_{us}$ are small, the
amplitudes for $\bskk$ and $\bskkneut$ are approximately equal in
the SM. However, this need not be the case for NP.

In order to determine the amplitudes for $\bskk$ and $\bskkneut$, we
need to know $P^{s\pm}$, $P^{s0}$, $T^{s\pm}$ and $T^{s0}$. These
can be obtained by considering $\bd$ decays. One can use $\bdpipi$
decays in order to determine the hadronic parameters
\cite{BLMV,LM,LMV,all1,all2}. However, it was noted in
Ref.~\cite{DMV} that $\bdKK$ provides smaller errors on the
predictions, and that the inclusion of some (quite generic) sign
predictions from $\bdpipi$ decays leads to a greater restriction on
the SM ranges.

The argument leading to the determination of the $\bdKK$ parameters
is as follows. The amplitude for this decay can be written
\beq {\cal A}(\bdKK) \simeq V_{ub}^* V_{ud} T^{d0} + V_{cb}^* V_{cd}
P^{d0} ~. \eeq
There are thus three unknown quantities to be determined: the
magnitudes of $P^{d0}$ and $T^{d0}$, and their relative strong
phase. Three pieces of information are therefore needed. One piece
of information comes from the measurement of the $\bdKK$ branching
ratio: $BR(\bdKK) = (0.96\pm 0.25) \times 10^{-6}$
\cite{brd}\footnote{While completing this work, this measurement has
been updated by BABAR in Ref.~\cite{Aubert:2006gm} to $BR(\bdKK) =
(1.08\pm 0.30) \times 10^{-6}$. However, in this work we prefer to
stick to the quoted value  more near to the HFAG value from
ICHEP06}.

A second piece of information comes from QCD factorization (QCDf)
\cite{QCDf1,QCDf2}. In QCDf the various hadronic quantities can be
calculated using a systematic expansion in $1/m_b$. However, a
potential problem arises when one encounters endpoint infrared (IR)
divergences in higher-order power-suppressed terms.  Their
evaluation thus requires an arbitrary IR cutoff, and they may be
enhanced numerically (for a given IR cutoff). The key observation
\cite{DMV} is that the difference $\Delta_d \equiv T^{d0} - P^{d0}$
is free of these IR divergences and can be calculated fairly
accurately within QCDf:
\beq \Delta_d = (1.09\pm 0.43) \times 10^{-7} + i (-3.02 \pm 0.97)
\times 10^{-7} ~{\rm GeV} ~. \label{Deltad} \eeq
Note that the values of the real and imaginary pieces of $\Delta_d$
can be affected by a global phase transformation, so that only the
modulus $|\Delta_d|$ is physical. This provides the second
constraint on the hadronic parameters of $\bdKK$.

Finally, the authors of Ref.~\cite{DMV} find that only values $-0.2
\le A_{dir}^{d0} \le 0.2$ are consistent with the measured value of
$BR(\bdKK)$ and the theoretical value of $\Delta_d$. This is the
third piece of information.

Using the values of the branching ratio and $|\Delta_d|$, as well as
the allowed range for $A_{dir}^{d0}$, one can obtain the moduli and
relative strong phase of the hadronic parameters in $\bdKK$:
\bea |T^{d0}| = (1.1 \pm 0.8) \times 10^{-6}~{\rm GeV} & , &
|P^{d0}/T^{d0}|
= 1.2 \pm 0.2 ~, \nn\\
{\rm arg}(P^{d0}/T^{d0}) & = & (-1.6 \pm 6.5)^{\circ} ~. \eea

In fact, there is a twofold discrete ambiguity in determining these
quantities, but the authors of Ref.~\cite{DMV} argue that only one
solution is physical. The argument uses the two methods involving
the decays $\bdKK$ and $\bdpipi$. It was shown in Ref.~\cite{DMV}
that the second solution (the unphysical one) requires a large
U-spin violation (in the phase) and, moreover, it predicts
$A_{dir}^{s\pm}<0$. This is clearly in contradiction with the
prediction for the sign of $A_{dir}^{s\pm}$ using $\bdpipi$, as can
be seen in Table 1 of Ref.~\cite{BLMV}. This resolves the two-fold
ambiguity.

In addition, it was pointed out that there exists a strong
anticorrelation between the signs of $A_{mix}^{s\pm}$ and
$A_{dir}^{d0}$, as can be seen in Table 1 of Ref.~\cite{DMV}. On the
other hand, using the method of $\bdpipi$ one can see that the
predicted value for $A_{mix}^{s\pm}$ is always negative \cite{BLMV}.
Putting together both methods, one finds an important restriction on
$A_{dir}^{d0}$ that prefers positive values. This affects and
improves all predictions, shown in Table 1 of Ref.~\cite{DMV} (only
the lower half of this table should be taken).

Recently, a measurement was reported by the BABAR collaboration
\cite{Aubert:2006gm}:
\beq A_{dir}(B_d \to K^0_s K^0_s)= -0.40\pm 0.41 \eeq
This very preliminary measurement, although still quite uncertain,
seems to point towards negative values for $A_{dir}^{d0}$, but it is
compatible with a small positive asymmetry. Note, however, that the
measured central value for $A_{mix}^{d0}$ is much larger than 1.
Thus, one should really take these numbers only as a proof that they
can be measured. Here we present the results for the range $-0.2 \le
A_{dir}^{d0} \le +0.2$ (this includes the non-preferred negative
region).

{}From these, one can now compute the parameters of the $\bskkneut$
amplitude, taking into account SU(3) breaking \cite{DMV}.
Factorizable SU(3)-breaking corrections are introduced by means of
$P^{s0} = f P^{d0}$ and $T^{s0} = f T^{d0}$, where the factor $f$ is
defined as \cite{QCDf2}
\beq f = {M_{\bs}^2 F_0^{\bs \to \kbar}(0) \over M_{\bd}^2 F_0^{\bd
\to \kbar}(0)} = 0.94 \pm 0.20 ~. \eeq
and the input values are taken from Refs.~\cite{pdg,QCDf2}. This
parameter can be calculated on the lattice.  Other sources of SU(3)
breaking which are suppressed by 1/$m_b$ originate from
hard-spectator scattering (differences in the distribution
amplitudes of $\bd$ and $\bs$ ) and weak annihilation (diagrams in
which the gluon emission comes from the spectator quark). All
effects are computed within QCDf (we assume that QCDf gives at least
the right order of magnitude), which gives the following bounds
\cite{DMV}:
\bea
|P^{s0}/(f P^{d0})-1| & \leq & 3\% ~, \nn\\
|T^{s0}/(f T^{d0})-1| & \leq & 3\% ~. \eea
Note that large final-state-interaction SU(3)-breaking effects are
possible. However, these are common to both $B_{d,s}^0 \rightarrow
K^0 \kbar$ decays and, consequently, they cancel in relating the two
modes. Thus the parameters of the amplitude for $\bskkneut$ can be
established (with errors), and the branching ratio and CP
asymmetries determined. These numbers are given below.

The decay $\bskk$ is somewhat more complicated, as a combination of
U-spin and isospin is required to connect $\bdKK$ to $\bskk$. The
relations between those hadronic parameters are given in
Ref.~\cite{DMV}, including SU(3)-breaking corrections evaluated
within QCDf. The bounds are
\bea
|P^{s\pm}/(f P^{d0})-1| & \leq & 2\% ~, \nn\\
|T^{s\pm}/(A_{KK}^{s} {\bar \alpha_1})-1-T^{d0}/(A^{d}_{KK} {\bar
  \alpha_1})| & \leq & 4\% ~,
\eea
where $A_{KK}^s$ and $\bar\alpha_1$ are additional hadronic
parameters that can be estimated in QCDf. However, it is also noted
that since the $T^{s\pm}$ term is CKM suppressed, any uncertainties
in the determination of $A_{KK}^s$ and $\bar\alpha_1$ affect the
branching ratio and CP asymmetries of $\bskk$ only marginally.

Putting all this together, the SM predictions for the branching
ratios and CP asymmetries in $\bskk$ and $\bskkneut$ can be
obtained.  If one conservatively takes all values of $A_{dir}^{d0}$
between $-0.2$ and $0.2$, the prediction for the branching ratio is
\cite{DMV}:
\bea BR(\bskkneut) & = & (18\pm 7 \pm 4 \pm 2)
\times 10^{-6} ~, \nn\\
BR(\bskk) & = & (20\pm 8 \pm 4 \pm 2) \times 10^{-6} ~. \eea
But if only  values of $A_{dir}^{d0}\geq 0$ (up to 0.2) are taken,
the prediction becomes:
\bea BR(\bskkneut) & = & (18\pm 7 \pm 4 \pm 2)
\times 10^{-6} ~, \nn\\
BR(\bskk) & = & (17 \pm 6 \pm 3 \pm 2) \times 10^{-6} ~. \eea
The first error reflects the uncertainty in the QCDf estimates of
$\Delta_d$ and $\bar\alpha_1$, as well as in $BR(\bdKK)$ (this is
the largest uncertainty). The second error corresponds to the
uncertainty in $f$ (SU(3) breaking). The third error introduces a
rough estimate of non-enhanced $1/m_b$-suppressed contributions.

\begin{figure}
\begin{center}
\psfrag{Adirs0}{\tiny \hspace{-1cm} $\stackrel{A_{\rm
dir}(\bskkneut)\times 10^{2}}{}$} \psfrag{Amixs0}{\tiny
\hspace{-1cm} $\stackrel{A_{\rm mix}(\bskkneut)\times 10^{2}}{}$}
\psfrag{Adirspm}{\tiny \hspace{-0.8cm} $\stackrel{A_{\rm
dir}(\bskk)}{}$} \psfrag{Amixspm}{\tiny \hspace{-0.8cm}
$\stackrel{A_{\rm mix}(\bskk)}{}$} \psfrag{Adird0}{}
\includegraphics[width=13cm]{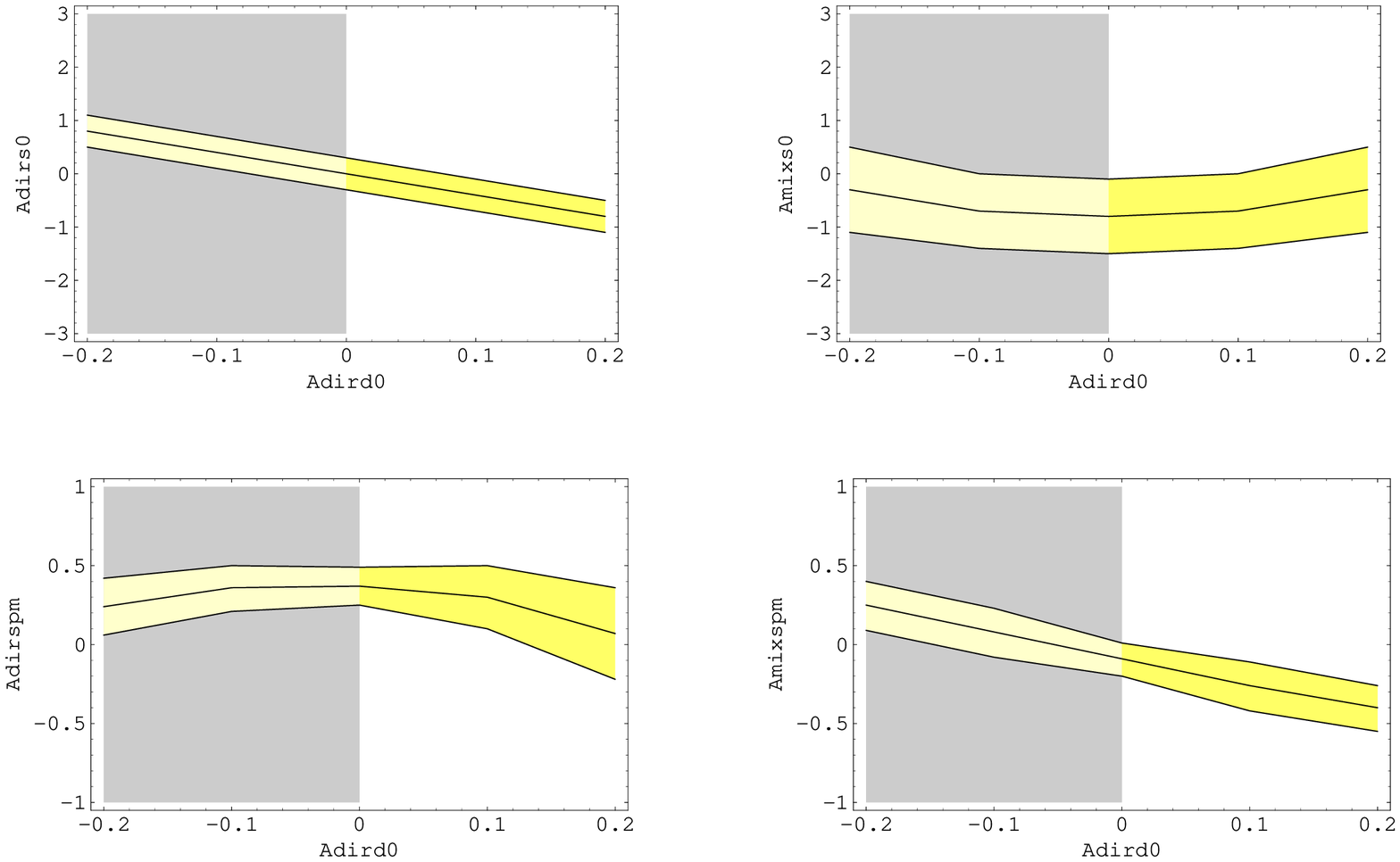}
\Text(-100,2)[lb]{\tiny $A_{\rm dir}(\bdKK)$}
\Text(-290,2)[lb]{\tiny $A_{\rm dir}(\bdKK)$}
\Text(-290,120)[lb]{\tiny $A_{\rm dir}(\bdKK)$}
\Text(-100,120)[lb]{\tiny $A_{\rm dir}(\bdKK)$}
\end{center}
\caption{SM predictions for the CP asymmetries in $\bskkneut$ (up)
and $\bskk$ (down) as a function of $A_{\rm dir}(\bdKK)$. As
explained in the text, the preferred range is the non-shadowed half
of the plots [$A_{\rm dir}(\bdKK)\geq 0$].} \label{plotACPs}
\end{figure}

The predicted ranges for the CP asymmetries of $\bskk$ and
$\bskkneut$ within the SM \cite{DMV} are illustrated in
Fig.\ref{plotACPs}.
The conservative predictions are \cite{DMV}:
\bea -0.011  \leq & A_{dir}^{s0} & \leq 0.011 ~, \nn \\
-0.015  \leq & A_{mix}^{s0} & \leq 0.005 ~, \nn \\
-0.22 \leq & A_{dir}^{s\pm} & \leq 0.49  ~, \nn\\
-0.55 \leq & A_{mix}^{s\pm} & \leq 0.40  ~. \eea
Or, considering only positive values of $A_{dir}^{d0}$:
\bea -0.011  \leq & A_{dir}^{s0} & \leq
0.003  ~, \nn \\
-0.015   \leq & A_{mix}^{s0} & \leq 0.005 ~, \nn \\
-0.22 \leq & A_{dir}^{s\pm} & \leq 0.49  ~, \nn\\
-0.55 \leq & A_{mix}^{s\pm} & \leq 0.02  ~. \eea
As expected, the CP asymmetries for $\bskkneut$ are predicted to be
very small in the SM, with the prediction that $|A_{dir}^{s0}|$
should be equal or less than 1\% and $|A_{mix}^{s0}|$ should be less
than 2\%.

\section{New Physics}

At present, there are many measurements of $B$ decays and several
discrepancies with the SM have appeared. For example, the CP
asymmetry in $\btos q{\bar q}$ modes ($q=u,d,s$) is found to differ
from that in ${\bar b} \to {\bar c} c {\bar s}$ decays by
2.6$\sigma$ (they are expected to be approximately equal in the SM)
\cite{HFAG,Zhou}. In addition, some $B\to\pi K$ measurements
disagree with SM expectations \cite{Baekandquim}, although the
so-called $B \to \pi K$ puzzle \cite{all2} has been reduced
\cite{puzzle,puzzle2}. One also sees a discrepancy with the SM in
triple-product asymmetries in $B \to \phi K^*$
\cite{BVVTP,phiKstarTP}, and in the polarization measurements of $B
\to \phi K^*$ \cite{BphiKstar_exp,BphiKstarNP, BphiKstarSM} and
$B\to \rho K^*$ \cite{BrhoKstar_exp,BrhoKstar}. Although these
discrepancies are not yet statistically significant, there is a
unifying similarity: they all point to new physics in $\btos$
transitions. We will therefore follow this indication and assume
that the NP appears in $\btos$ decays but does not affect $\btod$
decays. That is, $\bskk$ and $\bskkneut$ can be influenced by the
NP, but $\bdKK$ \cite{robert} is not.

There are many NP operators which can contribute to a given $B$
decay. However, in Ref.~\cite{DL}, it was observed that the matrix
elements of NP operators carrying non-negligible strong phases are
necessarily suppressed with respect to the SM contribution. (Note
that each NP contribution can in principle have a different strong
phase.) As a result, the relevant NP matrix elements can be
combined, and a given $B$ decay thus receives a single NP
contribution.

The amplitude for the decay $\bskk$ can therefore be written
\beq {\cal A}(\bskk) = {\cal A}^{s\pm}_{\sss SM} + \ANPu e^{i
\Phi_u} ~. \label{AK+K-} \eeq
In the above, ${\cal A}^{s\pm}_{\sss SM}$ contains both weak and
strong phases, but we have separated out the weak phase ($\Phi_u$)
of the NP contribution. (Note: the NP strong phase is zero, so that
$\ANPu$ is real.)  The name of the NP parameters reflects the fact
that this decay is ${\bar b} \to {\bar s} u {\bar u}$ at the quark
level.

Similarly, the amplitude for $\bskkneut$ (quark level: ${\bar b} \to
{\bar s} d {\bar d}$) is
\beq {\cal A}(\bskkneut) = {\cal A}^{s0}_{\sss SM} + \ANPd e^{i
\Phi_d} ~. \label{AK0K0} \eeq
If the NP conserves isospin, we have $\ANPu = \ANPd$ and $\Phi_u =
\Phi_d$, but in general this need not be the case.

One can make experimental measurements of $\bskk$, obtaining the
branching ratio and the direct and mixing-induced CP asymmetries.
Assuming that ${\cal A}^{s\pm}_{\sss SM}$ is known from $\bdKK$
decays (as in Sec.~2), these three measurements allow one to extract
$\ANPu$ and $\Phi_u$, as well as the relative strong phase between
the SM and NP amplitudes. One can proceed similarly for $\bskkneut$.
In this way one can {\it measure} all the NP parameters \cite{LMV}.

In the following section, we compute the SUSY contributions to
$\ANPu$, $\Phi_u$, $\ANPd$ and $\Phi_d$, and the resultant branching
ratios and CP asymmetries for $\bskk$ and $\bskkneut$.

\section{\boldmath SUSY Predictions for $\bskk$ and $\bskkneut$}

As mentioned above, the relevant SUSY contributions to ${\bar b} \to
{\bar s} q {\bar q}$ transitions come from squark-gluino box and
penguin diagrams. We follow the procedure outlined in
Ref.~\cite{BLMV}, which is based on the work of Grossman, Neubert
and Kagan \cite{trojan}.  We refer the reader to these references
for further details.

As shown in Ref.~\cite{Baek:2004}, the most natural solution to the
$B \to \pi K$ puzzle is the introduction of isospin-breaking NP
amplitudes. The isospin-breaking effect is more naturally realized
in the present scenario, which allows large up-down squark mass
splittings, than in the more popular mass insertion (MI)
approximation. Indeed, since we are working in a scenario with
near-maximal mixing between bottom and strange squarks, the squark
MI approximation is not adequate.  Note that the dangerous ${\bar s}
\to {\bar d}$ and $\btod$ flavour-changing neutral currents are not
generated in this scenario due to the assumption of vanishing
$(1,2)$ and $(1,3)$ components in the scalar down-type squark mass
matrix \cite{trojan}.

In this scenario, the SUSY contributions to the Wilson coefficients
depend on the following parameters: the masses of the squarks and
gluino, two mixing angles $\theta_{L,R}$ and two weak phases
$\delta_{L,R}$. These angles parametrize the rotation matrices that
diagonalize the left- and right-handed squark mass matrices. The
expressions for the NP amplitudes $\ANPu e^{\Phi_u}$ and $\ANPd
e^{\Phi_d}$ in terms of these parameters are obtained in complete
analogy with Ref. \cite{BLMV}. To be specific, we write the
expressions for these amplitudes in terms of the Wilson coefficients
(denoted by $\bar{c}_i^q$ and $\bar{C}_{8g}^{\rm eff}$):
\begin{eqnarray}
\mathcal{A}^u e^{i\Phi_ u}&=&\bra{K^+K^-}\mathcal{H}^{\rm NP}_{\rm eff}\ket{B_s^0}\nn\\
&=&\frac{G_F}{\sqrt{2}} \Big[
-\chi(\frac{1}{3}\bar{c}_1^u+\bar{c}_2^u)-\frac{1}{3}(\bar{c}_3^u-\bar{c}_6^u)-(\bar{c}_4^u-\bar{c}_5^u)
-\lambda_t \frac{2\alpha_s}{3\pi}\bar{C}_{8g}^{\rm eff}\big( 1+\frac{\chi}{3} \big) \Big]A\nn\\
\mathcal{A}^d e^{i\Phi_ d}&=&\bra{K^0\bar{K}^0}\mathcal{H}^{\rm NP}_{\rm eff}\ket{B_s^0}\nn\\
&=&\frac{G_F}{\sqrt{2}} \Big[
-\chi(\frac{1}{3}\bar{c}_1^d+\bar{c}_2^d)-\frac{1}{3}(\bar{c}_3^d-\bar{c}_6^d)-(\bar{c}_4^d-\bar{c}_5^d)
-\lambda_t \frac{2\alpha_s}{3\pi}\bar{C}_{8g}^{\rm eff}\big(
1+\frac{\chi}{3} \big) \Big]A\qquad\quad
\end{eqnarray}
where $\chi\simeq 1.18$ and
\beq
A=\bra{\bar{K}^0}(\bar{b}d)_{V+A}\ket{B_s^0}\bra{K^0}(\bar{d}s)_{V+A}\ket{0}=
i (m_B^2-m_K^2) f_K F^{B_s\to K}\simeq 1.42 ~{\rm GeV}^3 \eeq
For the numerical values we take \cite{pdg,QCDf2} $F^{B_s\to
K}=0.31$ and $f_K=0.1598\,{\rm GeV}$. The explicit expressions for
the Wilson coefficients ${\bar c}^{d,u}_i$ (which includes both L
and R mixing contributions) can be found in the appendix of
Ref.~\cite{BLMV}, where some small typos in Ref.~\cite{trojan} were
corrected.

In order to obtain the NP amplitudes, we must evaluate the Wilson
coefficients within SUSY. We consider the following values and
ranges for the SUSY parameters:
\begin{itemize}

\item
$m_{\tilde{u}_L}=m_{\tilde{d}_{L,R}}=m_{\tilde{b}_{L,R}}=m_{\tilde{g}}=
250 ~{\rm GeV}$

\item
$250 ~{\rm GeV} < m_{\tilde{u}_R}, m_{\tilde{s}_{R,L}} < 1000 ~{\rm
GeV}$

\item $-\pi < \delta_{L,R} < \pi$

\item $-\pi/4 < \theta_{L,R} < \pi/4$

\end{itemize}
The SM inputs are the same as those used in Ref.~\cite{DMV}. The
Wilson coefficients are sensitive to the $\tilde{s}-\tilde{b}$ mass
splitting. They vanish for $m_{\tilde{s}}=m_{\tilde{b}}$ and grow
when the splitting is large. We therefore expect these contributions
to be most important for large values of $m_{\tilde{s}}$ (keeping
$m_{\tilde{b}}$ fixed). In the same way, NP effects in ${\bar b} \to
{\bar d} q {\bar q}$ transitions depend on the difference
$m_{\tilde{d}}-m_{\tilde{b}}$. By setting
$m_{\tilde{d}}=m_{\tilde{b}}$ we ensure that $\btod$ decays get no
such contributions, which is consistent with the discussion in
Sec.~3. A difference between $\ANPu e^{\Phi_u}$ and $\ANPd
e^{\Phi_d}$ is only possible in the presence of a nonzero
$\tilde{u}-\tilde{d}$ mass splitting. Without it there are no
contributions to isospin-violating operators. However, this mass
splitting must be very small in the left-handed sector due to
$SU(2)_L$ invariance.  We therefore set
$m_{\tilde{u}_L}=m_{\tilde{d}_L}$, but allow for a significant mass
splitting in the right-handed sector.

\begin{figure}
\begin{center}
\psfrag{IAu}{\footnotesize \hspace{-1.2cm} $\rm{Im}(\ANPu e^{i
\Phi_u})\times 10^{9}$} \psfrag{IAd}{\footnotesize \hspace{-1.2cm}
$\rm{Im}(\ANPd e^{i \Phi_d})\times 10^{9}$} \psfrag{RAu}{}
\psfrag{RAd}{}
\includegraphics[height=7cm]{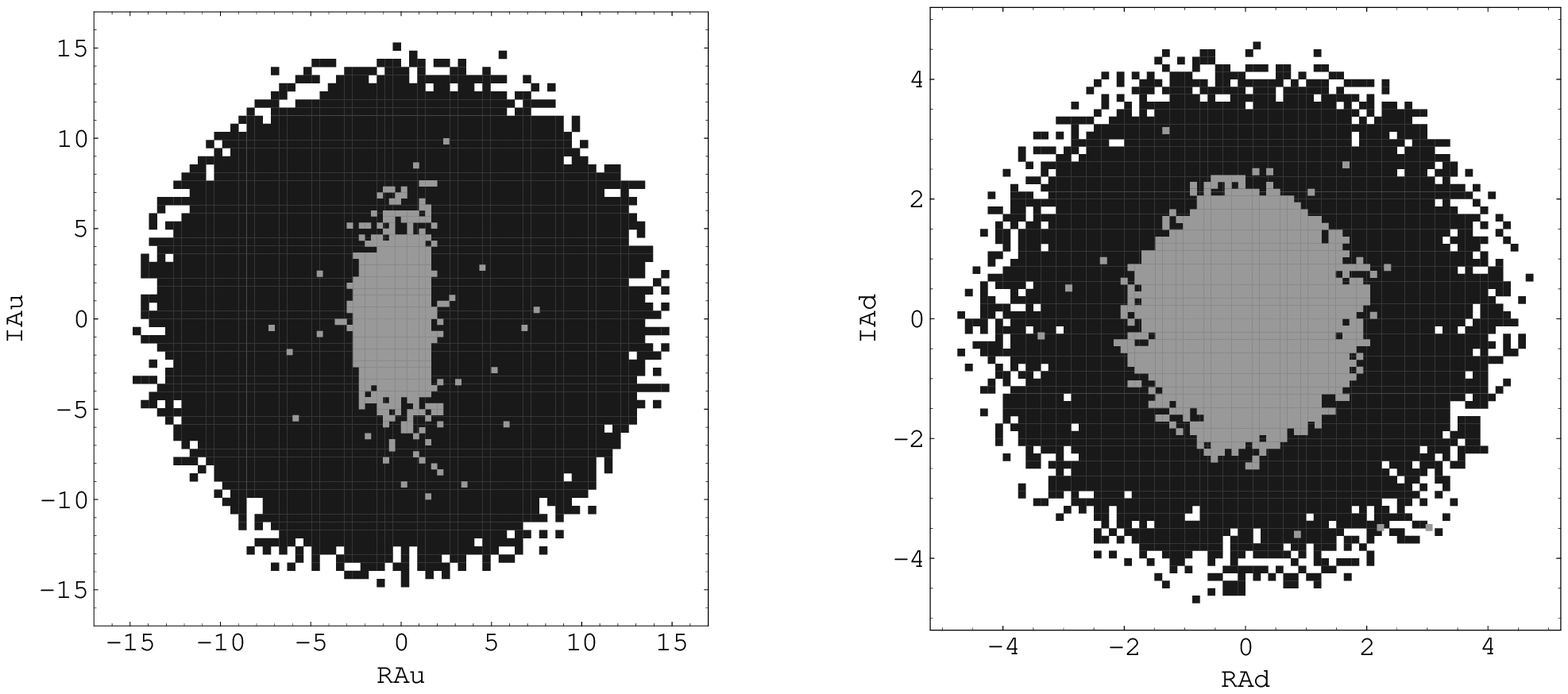}
\Text(-135,0)[lb]{\footnotesize $\rm{Re}(\ANPd e^{i \Phi_d})\times
10^{9}$} \Text(-355,0)[lb]{\footnotesize $\rm{Re}(\ANPu e^{i
\Phi_u})\times 10^{9}$}
\end{center}
\caption{\small SUSY contribution to the NP amplitudes $\ANPu e^{i
\Phi_u}$ (left) and $\ANPd e^{i \Phi_d}$ (right) in the scenario
with simultaneous LL and RR mixings.  The dark regions correspond to
the variation of the SUSY parameters over the considered parameter
space. The light regions satisfy the experimental bounds, including
the recent measurement of $\Delta M_s$.} \label{plotANP}
\end{figure}

There are also constraints on the SUSY parameter space from other
processes that have been already measured. The constraints from the
decays $B\to \pi K$ and $B \to X_s \gamma$ are described in
Ref.~\cite{BLMV}. In particular, we take $BR(B\to
X_s\gamma)=(3.55\pm 0.26)\times 10^{-4}$ \cite{HFAG}, with an
increased error to cover the various theoretical
uncertainties\footnote{Interestingly, the latest NNLO calculations
in the SM show that $BR(B\to X_s\gamma)$ (SM) is a little lower than
the experimental average \cite{bsgamma_NNLO}.}. Most importantly,
one must take into account the recent measurement of $\Delta M_s$,
which was not included in the previous analysis.  The latest
measurement \cite{Bsmix}, together with the SM fit
\cite{CKMfit,Bona}, gives\footnote{Note that here we take the
largest fit result for the SM prediction \cite{CKMfit} because it
falls within the second prediction of Ref.~\cite{Bona}. Moreover,
the value for ${\Delta M_s^{\rm \sss SM}}$ taken here differs from
that used for the bound in Ref.~\cite{BLMV} that was based on
Ref.~\cite{PRS}.}
\beq \bigg(\frac{\Delta M_s}{\Delta M_s^{\rm \sss
SM}}\bigg)_{exp}=(0.81\pm 0.19)~{\rm ps}^{-1} \eeq
The constraints from all these measurements have been included in
our analysis. Other traditionally-important constraints like $\bs\to
\mu^+\mu^-$ are very sensitive to other SUSY parameters, mostly
$\tan{\beta}$ and $m_A$. However, for small values of $\tan{\beta}$
and values for $m_A$ above 200 GeV, they have no effect on our
allowed region to SUSY (Fig.~\ref{plotANP}).

Taking into account the various constraints, the contributions from
LL and RR mixing have been analyzed. $\Delta M_s$ is the strongest
constraint, and it is the relevant one when considering only LL or
RR mixing separately. In particular, it has a large impact on the
phases $\Phi_u$ and $\Phi_d$. In the case of $\bskk$, LL mixing
gives the largest contribution to the amplitude, more than twice
that of RR. However, in the case of $\bskkneut$ both contributions
are similar in size.

When both LL and RR mixings are allowed simultaneously, the
constraints on the SUSY parameter space are changed. In this case
new operators for $\bs$--$\bsbar$ mixing are generated, so that the
effect is not simply the combination of the two separate
contributions (for instance, see Ref.~\cite{Baek2}). We find that
(i) now $BR(B\to X_s\gamma)$ is also important, not only $\Delta
M_s$, and (ii) the global effect of the constraints is weaker.  The
upshot is that there is a certain enhancement of the NP amplitudes
when both LL and RR mixings are combined. In particular, the weak
phases $\Phi_u$ and $\Phi_d$ are not so strongly constrained as when
either LL or RR mixing is taken to vanish.

In Fig.~\ref{plotANP} we show the allowed ranges for $\ANPu e^{i
\Phi_u}$ and $\ANPd e^{i \Phi_d}$ in the scenario with simultaneous
LL and RR mixings. The dark regions correspond to the values that
these amplitudes take when varying the parameter space over the
initial ranges. The light regions show how these values are reduced
by the existing experimental constraints mentioned above. There are
two important remarks. First, we see that the above constraints do
indeed greatly reduce the allowed SUSY parameter space. Second, even
so, the effect on $\ANPu e^{i \Phi_u}$ and $\ANPd e^{i \Phi_d}$ can
be significant.

At this stage we can identify what are the effects of the various
constraints in reducing the SUSY parameter space. The bound from $B
\to X_s \gamma$ affects only the left-handed sector. In particular,
for large $m_{\tilde{s}}$, the regions with $|\theta_L| \gtrsim
10^\circ$, $|\delta_L| \lesssim 60^\circ$ and $|\theta_L| \gtrsim
10^\circ$, $|\delta_L-\pi| \lesssim 60^\circ$ are excluded.  The
bound from $\Delta M_s$ is much stronger: when $m_{\tilde{s}}\gtrsim
400~{\rm GeV}$, any values of $|\theta_{L,R}|\gtrsim 5^\circ$ are
excluded, as well as those regions in which
$\delta_L+\delta_R\approx -3\pi/2,-\pi/2,\pi/2,3\pi/2$.\footnote{For
further details of the $\Delta M_s$ constraint on this parameter
space, see Ref.~\cite{Baek2}.} After these bounds are imposed on the
parameter space, the constraints from $B\to \pi K$ have very little
effect on the regions in Fig.~\ref{plotANP}.

Note that the allowed region for $\ANPu$ is much larger than that
for $\ANPd$, by approximately a factor of 3. In the isospin limit,
these should be equal, so this factor of 3 is a measure of isospin
breaking in this NP scenario. In particular, for
$m_{\tilde{u}_R}={\rm 250~GeV}$ (zero $\tilde{u}_R$-$\tilde{d}_R$
mass splitting), the values of $\ANPu$ reduce to those for $\ANPd$.

\begin{figure}
\begin{center}
\psfrag{BR}{\hspace{-1.5cm} \tiny $\stackrel{BR(\bskk)\times
10^6}{}$} \psfrag{Amix}{\hspace{-1cm} \tiny $\stackrel{A_{\rm
mix}(\bskk)}{}$} \psfrag{Adir}{}
\includegraphics[width=15.5cm]{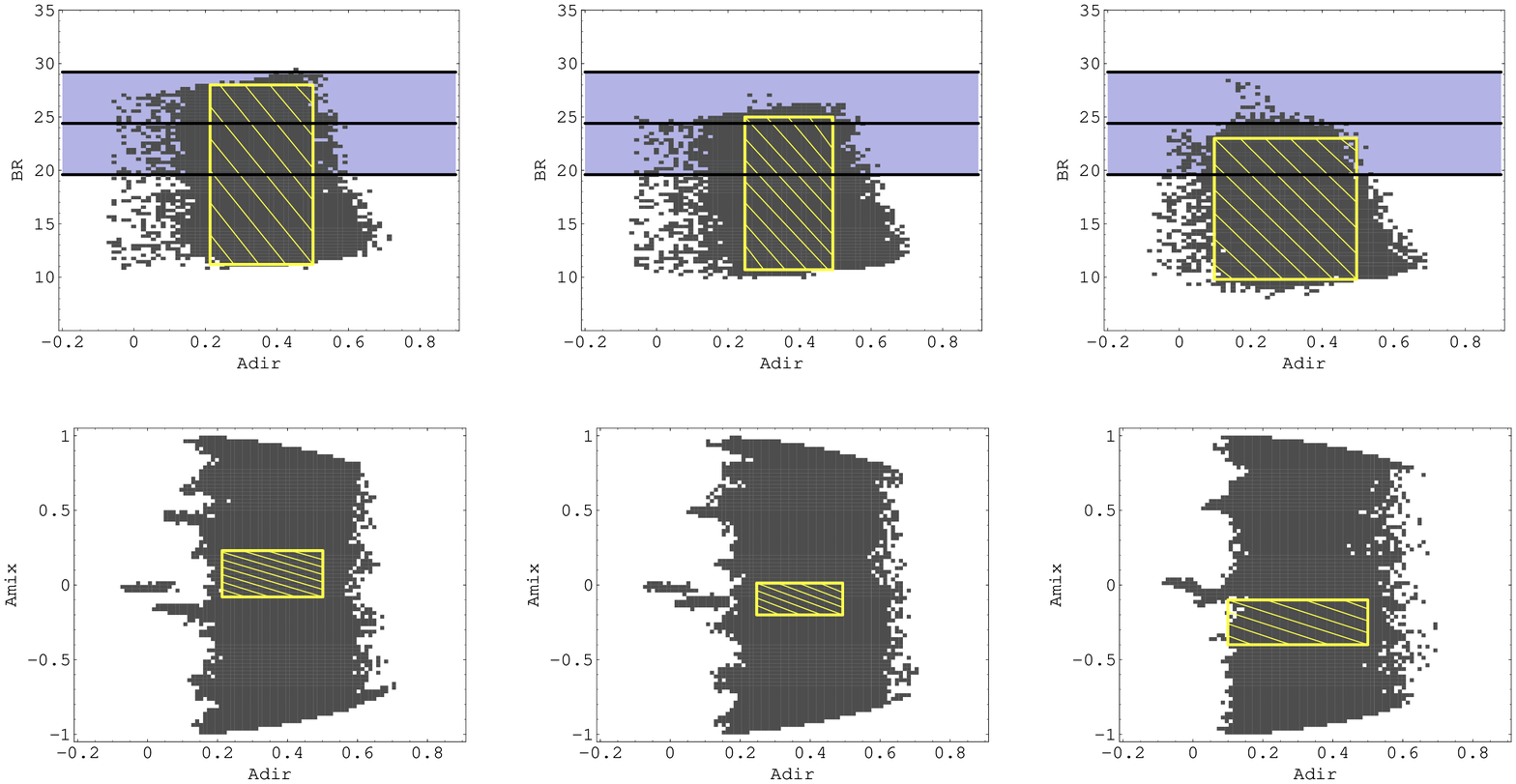}
\Text(-100,5)[lb]{\tiny $A_{\rm dir}(\bskk)$}
\Text(-245,5)[lb]{\tiny $A_{\rm dir}(\bskk)$}
\Text(-390,5)[lb]{\tiny $A_{\rm dir}(\bskk)$}
\Text(-100,120)[lb]{\tiny $A_{\rm dir}(\bskk)$}
\Text(-245,120)[lb]{\tiny $A_{\rm dir}(\bskk)$}
\Text(-390,120)[lb]{\tiny $A_{\rm dir}(\bskk)$}
\Text(-360,-15)[lb]{\small (a)} \Text(-215,-15)[lb]{\small (b)}
\Text(-70,-15)[lb]{\small (c)}
\end{center}
\caption{\small Predictions, in the form of scatter plots, for the
correlations between $BR(\bskk)-A_{\rm dir}(\bskk)$ (up) and $A_{\rm
mix}(\bskk)-A_{\rm dir}(\bskk)$ (down) in the presence of SUSY, for
a) $A_{dir}^{d0}=-0.1$, (b) $A_{dir}^{d0}=0$ and (c)
$A_{dir}^{d0}=0.1$. The dashed rectangles correspond to the SM
predictions. The horizontal band shows the experimental value for
$BR(\bskk)$ at  $1\sigma$.} \label{plotBK+K-}
\end{figure}

We now examine the effect of these contributions on the observables.
By adding the SUSY contributions to the SM amplitudes as in
Eqs.~(\ref{AK+K-}) and (\ref{AK0K0}), it is possible to compute the
branching ratio and the CP asymmetries in the presence of SUSY.
Fig.~\ref{plotBK+K-} shows the allowed values for the $\bskk$
observables, for three different values of $A_{dir}^{d0}$, compared
with the predictions of the SM and with the recent experimental
value for the $\bskk$ branching ratio reported by the CDF
collaboration \cite{CDF} [Eq.~(\ref{bskk_CDF})]. The agreement
between the CDF measurement and the prediction of the SM in
Ref.~\cite{DMV} erases any discrepancy between experiment and the
SM. This branching ratio will now be an important future constraint.
The branching ratio within SUSY should not deviate much from the SM
prediction so as not to generate any disagreement with data. Indeed,
Fig.~\ref{plotBK+K-} shows that the impact of SUSY on the branching
ratio of $\bskk$ is practically negligible.  Interestingly, for
positive values of $A_{dir}^{d0}$ (preferred region), the SM
predicts a smaller value for $BR(\bskk)$, but it is now compatible
with the new data. Still, it is in this case that SUSY shows a
larger deviation in the correct direction.

A completely different picture arises for the CP asymmetries. The
results for the direct CP asymmetry reveal that SUSY can have an
impact. This is not surprising: SUSY introduces a term in the total
amplitude which is of the same order of magnitude as that of the SM
and carries a weak phase that is not constrained.  The
mixing-induced CP asymmetry gets affected in a more dramatic way.
The interpretation is that the SUSY contribution to the mixing angle
$\phi_s$ can be large (in fact it can take all values between $-\pi$
and $\pi$), while in the SM it is tiny: $\phi_s^{SM}\simeq
-2^\circ$. Any experimental measurement falling inside the dark area
in the plots, but outside the dashed rectangle, would not only
signal NP but clearly could be accommodated by supersymmetry.

\begin{figure}
\begin{center}
\psfrag{BR}{\hspace{-1.5cm} \tiny $\stackrel{BR(\bskkneut)\times
10^6}{}$} \psfrag{Amix}{\hspace{-1cm} \tiny $\stackrel{A_{\rm
mix}(\bskkneut)}{}$} \psfrag{Adir}{}
\includegraphics[width=15.5cm]{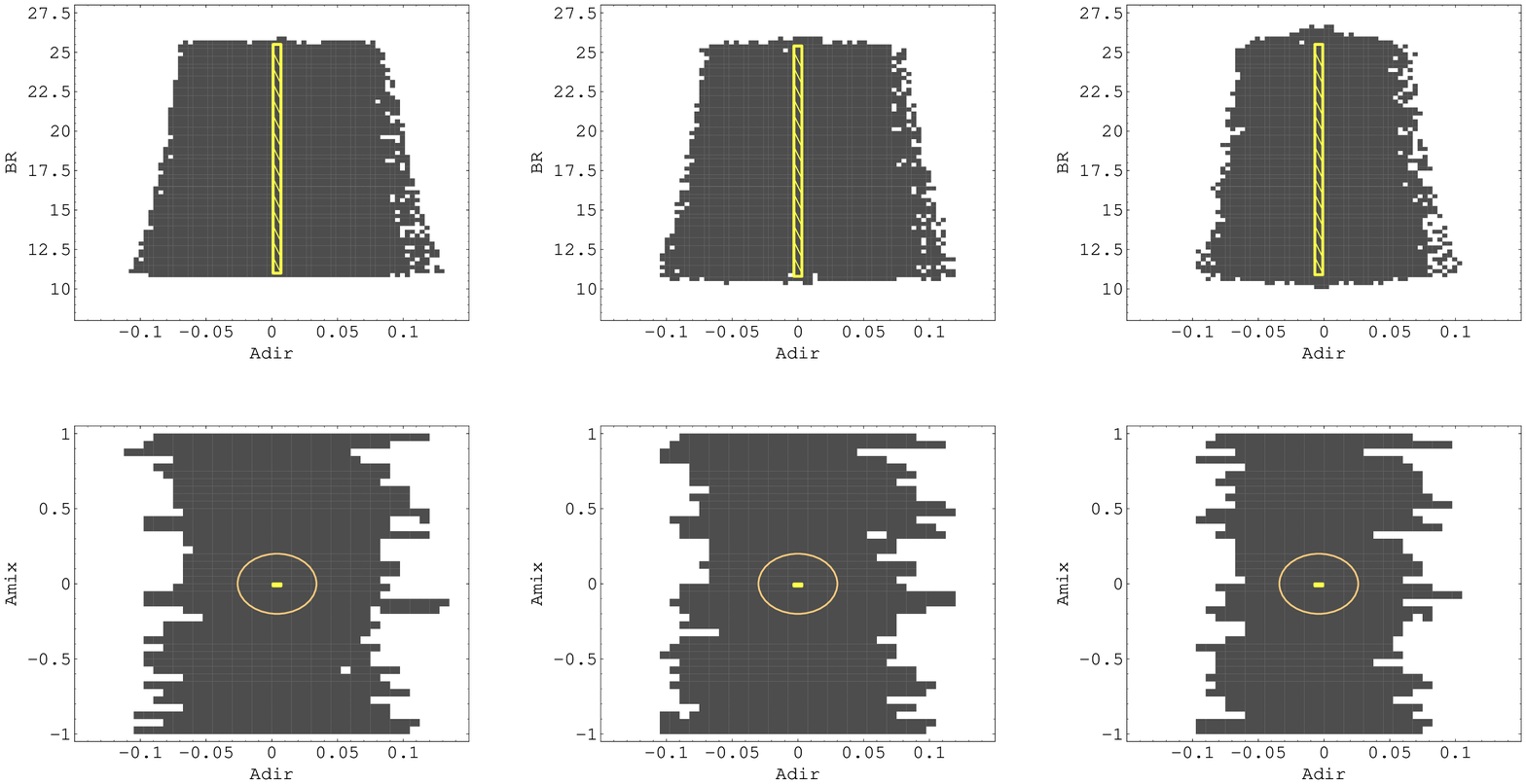}
\Text(-100,5)[lb]{\tiny $A_{\rm dir}(\bskkneut)$}
\Text(-245,5)[lb]{\tiny $A_{\rm dir}(\bskkneut)$}
\Text(-390,5)[lb]{\tiny $A_{\rm dir}(\bskkneut)$}
\Text(-100,120)[lb]{\tiny $A_{\rm dir}(\bskkneut)$}
\Text(-245,120)[lb]{\tiny $A_{\rm dir}(\bskkneut)$}
\Text(-390,120)[lb]{\tiny $A_{\rm dir}(\bskkneut)$}
\Text(-360,-15)[lb]{\small (a)} \Text(-215,-15)[lb]{\small (b)}
\Text(-70,-15)[lb]{\small (c)}
\end{center}
\caption{\small Predictions, in the form of scatter plots, for the
correlations between $BR(\bskkneut)-A_{\rm dir}(\bskkneut)$ (up) and
$A_{\rm mix}(\bskkneut)-A_{\rm dir}(\bskkneut)$ (down) in the
presence of SUSY, for (a) $A_{dir}^{d0}=-0.1$, (b) $A_{dir}^{d0}=0$
and (c) $A_{dir}^{d0}=0.1$. The dashed rectangles correspond to the
SM predictions. These are quite small in the three lower plots, so
they are indicated by a circle.} \label{plotBK0K0}
\end{figure}

Fig.~\ref{plotBK0K0} shows the results for $\bskkneut$. Although the
branching ratio is little changed in the presence of SUSY, the
enhancement of the CP asymmetries due to the inclusion of the SUSY
contributions is in this case even more important. The reason is
that, within the SM, the CP asymmetries are much smaller in
$\bskkneut$ than they are for $\bskk$, because of the absence of the
tree diagram. Thus the impact of SUSY is much greater. This is
evident by looking at the three lower plots in Fig.~\ref{plotBK0K0},
where the tiny rectangles corresponding to the SM predictions can
hardly be observed. We have drawn a circle around them to indicate
their position.

These are a good illustration of the general scenario discussed in
the introduction.  While the branching ratios in this case are
relatively insensitive to supersymmetry, the direct and
mixing-induced CP asymmetries of these decays are greatly affected.
Thus, these CP asymmetries are the observables to focus on in order
to observe NP, particularly SUSY, while the branching ratio of
$\bskk$ can become an important constraint on models beyond the SM
other than SUSY.

\section{Conclusions}

In this paper we have considered the branching ratios and CP
asymmetries for the decays $\bskk$ and $\bskkneut$ in a
supersymmetric (SUSY) model, focusing on the dominant gluino-squark
contributions \cite{trojan}. This analysis is an extension and
update of Ref.~\cite{BLMV}, which now includes the SUSY analysis of
both $\bs$ decay modes and allows for both LL and RR mixing. In
addition, the determination of the SM contributions from the decay
$\bd \to \pi^+\pi^-$ used in Ref.~\cite{BLMV} has been replaced by
the recently-proposed combination of $\bdKK$ and QCD factorization
of Ref.~\cite{DMV}. This allow us to obtain more precise
predictions.

We have included the constraints coming from $BR(B \to X_s \gamma)$,
$B \to \pi K$ and $\Delta M_s$, and we find the following results.
\begin{itemize}

\item The new-physics (NP) amplitudes are $\ANPu e^{i \Phi_u}$
($\bskk$) and $\ANPd e^{i\Phi_d}$ ($\bskkneut$). We find that both
can get significant contributions from SUSY. In the isospin limit,
these quantities are equal. However, our calculations show that, for
the region of parameters considered, in SUSY there can be a
difference of up to a factor of 3 between the NP amplitudes. This
indicates the possible level of isospin breaking in this type of
theory. In particular, in the SUSY model considered here, large
isospin violation is possible when there is large mass splitting in
$\widetilde{u}_R$-$\widetilde{d}_R$.

\item The branching ratio $BR(\bskk)$ is very little affected by
SUSY. At most, the SM prediction can be increased by 15\% for
$A_{dir}^{d0}=0.1$. In fact, SUSY can somewhat improve the already
good agreement between the SM prediction and the new precise CDF
measurement \cite{CDF}. The impact of SUSY on $BR(\bskkneut)$ is
even smaller, reflecting the reduced allowed region for $\ANPd
e^{i\Phi_d}$ as compared to $\ANPu e^{i \Phi_u}$.

\item The situation is very different for the CP asymmetries; the size
of the effect depends strongly on the decay and the type of
asymmetry. For $\bskk$, the direct CP asymmetry within SUSY is in
the range $-0.1 \lsim A_{dir}(\bskk)^{SUSY} \lsim 0.7$ for $-0.1
\leq A_{dir}^{d0} \leq 0.1$. Depending on the value of
$A_{dir}^{d0}$, it may be possible to disentangle the SUSY
contribution from that of the SM. This is due to the competition
between the tree and the NP amplitudes for each value of
$A_{dir}^{d0}$. As for $A_{mix}(\bskk)$, its value can vary all the
way from $-1$ to $+1$, signaling a large impact from SUSY.

\item Turning to $\bskkneut$, the CP asymmetries are particularly
promising. This decay is dominated by the penguin amplitude in the
SM, and so the direct CP asymmetry is strongly suppressed: it is
predicted to be at most of the order of 1\%. However, in the
presence of SUSY, the direct CP asymmetry can be 10 times larger.
The mixing-induced CP asymmetry is also predicted to be very small
in the SM. However, $A_{mix}(\bskkneut)^{SUSY}$ covers the entire
range, and so this asymmetry can be large in the presence of SUSY.

\end{itemize}

\bigskip
\noindent {\bf Acknowledgements}: \\ We thank G. Punzi for helpful
discussions. This work was financially supported by the Korea
Research Foundation Grant funded by the Korean Government (MOEHRD)
No. KRF-2005-070-C00030 (SB), NSERC of Canada (DL), and by
FPA2005-02211 (JM \& JV), PNL2005-41 (JM \& JV) and the Ramon y
Cajal
Program of Spain (JM).\\\\


\end{document}